\documentstyle[referee]{l-aa}
\input psfig.tex
\def\lsim{\lower.5ex\hbox{$\; \buildrel < \over \sim \;$}}
\def\gsim{\lower.5ex\hbox{$\; \buildrel > \over \sim \;$}}
\def \simeq{\lower.3ex\hbox{$\; \buildrel \sim \over - \;$}}
\def\ch{\lower-0.55ex\hbox{--}\kern-0.55em{\lower0.15ex\hbox{$h$}}}
\def\lh{\lower-0.55ex\hbox{--}\kern-0.55em{\lower0.15ex\hbox{$\lambda$}}}

.tfm

\begin{document}
\title{The effect of cooling on time dependent behaviour of accretion flows around black holes}
\author{Sandip K. Chakrabarti$^{1,2}$, K. Acharyya$^2$ and D. Molteni$^3$}
\institute {$^1$ S. N. Bose National Centre for Basic Sciences, Salt Lake, Kolkata 700098, India\\
$^2$ Centre for Space Physics, Chalantika 43, Garia Station Rd., Kolkata, 700084, India\\ 
$^3$ Dipartimento di Fisica e Tecnologie Relative, Viale delle Scienze, 90128 Palermo Italy\\
e-mails: chakraba@bose.res.in, space\_phys@vsnl.com, molteni@unipa.it}
\offprints{S. K. Chakrabarti {\it chakraba@boson.bose.res.in}}
\date{Received ; accepted , }
\maketitle
\markboth{  }{}

\begin{abstract}

We present the results of several numerical simulations of two dimensional 
axi-symmetric accretion flows around black holes using Smoothed Particle Hydrodynamics (SPH)
in the presence of cooling effects. We consider both stellar black holes and 
super-massive black holes. We observe that due to {\it both radial and vertical
oscillation} of shock waves in the accretion flow, the luminosity and average thermal energy 
content of the inner disk exhibit very interesting behaviour. When power 
density spectra are taken, quasi-periodic variabilities are seen at a few 
Hz and also occasionally at hundreds of Hz for stellar black holes. 
For super-massive black holes, the time scale of the oscillations
ranges from hours to weeks. The power density spectra have a flat 
top behavior with average {\it rms} amplitude of a few percent 
and a broken power-law behavior. The break frequency is generally found to be
close to the frequency where the shock oscillates.

\end{abstract}

\section{Introduction}

X-rays emitted by galactic black hole candidates are known to exhibit quasi-periodic variations (QPVs).
Today, there are almost a dozen confirmed stellar mass black hole candidates for which 
QPVs are regularly observed, some at normal frequencies ($0.1-10$Hz) and some with
frequencies of hundreds of Hz. While it is conceptually easier to explain such variations when there is a 
hard surface, as on a neutron star (e.g., Shaham, 1987), in black hole candidates they are
difficult to understand as there are no surfaces from which a perturbation could be reflected back
so as to produce a sustained oscillation of significant amplitude. 

In a very important work, Langer, Chanmugam and Shaviv (1981) first suggested that 
in the presence of cooling effects standing shocks on a white dwarf surface can 
oscillate and in fact the oscillation may also subside when 
cooling is increased (Chanmugam, Langer, and Shaviv, 1985). While a black hole does not have a 
hard surface like neutron stars or white dwarfs, the centrifugal force in accretion matter
close to a black hole can become strong enough to actually slow down matter and form 
centrifugal-pressure-supported axisymmetric standing shocks 
(Chakrabarti, 1989) depending on specific energy and angular momentum 
(this region will be referred to as CENBOL, centrifugal pressure dominated boundary layer). 
In the particle dynamics picture, the appearance of a discontinuity can be argued this way: the 
centrifugal force $F_{cen} \propto \lambda^2/r^3$ grows more rapidly 
(for constant or almost constant specific angular momentum $\lambda$)
than the gravitational force $F_{G} \propto GM/r^2$ and they become comparable at a distance of 
$\lambda \sim (GMr)^{1/2}$, where $M$ is the mass of the central compact object.
In fluid dynamics, because of the addition of thermal pressure  and/or radiation  pressure, 
such a balance takes place farther out where a shock may develop. 
The almost constancy of angular momentum comes into being
primarily because, as has been shown by extensive work done earlier (Chakrabarti, 1996a), 
the viscosity timescale $t_{visc} \sim r^2/\nu_s$ is much longer than the
the infall time scale $r^{3/2}/(GM)^{1/2}$, $\nu_s$ being the kinematic viscosity. 
It was shown by Molteni, Sponholz \& Chakrabarti (1996, hereafter referred 
to as MSC96) and Ryu, Chakrabarti, \& Molteni (1997, hereafter 
RCM 1997) that these shocks may also oscillate in the presence of cooling. 
These oscillations are explained to be due to a resonance between the dominant 
cooling time scale and the infall time scale, and the cooling could be due to 
thermal/non-thermal radiative effects (as in MSC96), or to
dynamical cooling due to outflows (as in RCM97) or both. This shock
forms in low-angular-momentum (in the sense mentioned), quasi-spherical
flows which may or may not surround a
Keplerian disk located in the equatorial plane. In case the latter disk is present
and the soft photons emitted are intercepted, inverse-Comptonized and re-emitted by the 
post-shock  region, as in the two-component advective flow model (Chakrabarti
\& Titarchuk, 1995; hereafter CT95), then, as the shock oscillates, 
the intensity of hard X-rays is also modulated at the same frequency. 

If our shock oscillation solution is correct, then a large number of simple 
predictions could be made: (a) the harder radiation
coming from the hot, post-shock region would primarily participate in QPVs while the softer
radiation would have less power. This was verified by Chakrabarti \& Manickam 
(2000, hereafter CM00) and Rao et al. (2000). (b) the stronger a
shock is, the stronger is the thermally driven outflow rate. Thus, 
the outflow from the post-shock region depends on the shock strength, 
i.e., the spectral states (Chakrabarti, 1999). 
This has also been tested by various observations (e.g., 
Dhawan et al., 2000). (c) In case of outflows, the electron density in 
the post-shock region is reduced and the Comptonized spectrum should be softened.
Chakrabarti et al. (2003) showed that 
indeed the presence/absence of outflows does change the spectral slope 
in the hard X-ray region. These observations vindicate that the
Comptonized photon, and the major part of the outflows both
come out of the post-shock region. Thus, a thorough study of 
the behavior of CENBOL, especially when it is time dependent, is needed.

In the present paper, we carry out numerical simulation of thick, 
inviscid, rotating axisymmetric flows, in the presence of 
cooling effects for a large region of the parameter space. 
The hot electrons in the CENBOL
reprocess soft photons and generally enhance their
energy before these photons escape. The degree by which the 
energy is transferred depends on the so-called 
{\it  Compton y parameter} (Rybicki \& Lightman, 1979),
$$
y=\tau_0 (4 \Theta + 16 \Theta^2) {\rm max(1,}\tau_0)
\eqno{(1)}
$$
where $\tau_0$ is the optical depth of the CENBOL, 
and $\Theta=kT_e/m_e c^2$ is dimensionless electron temperature,
($k$, $T_e$, $m_e$ and $c$ being the Boltzmann constant, electron temperature,
electron mass and the velocity of light respectively).
If $y<<1$, nothing much happens to the photons, and for $y>>1$
the saturated Comptonization causes a Wien distribution to form 
at a high energy. The most important case is where $y$ is around $1$. In this
latter case, the energy loss from electrons to photons becomes important. To calculate this
cooling effect in a time-dependent simulation is not easy, since ideally, at 
every time step, one has to carry out a Monte-Carlo simulation of the 
configuration of the flow. Here we use a shortcut to the process. 
To include the effect of Comptonization, we borrow the
results of a Monte Carlo simulation by Dermer, Liang \& Canfield (1991, hereafter DLC91)
where an enhancement factor $\eta$ 
(the ratio of up-scattered flux to injected flux) has been calculated 
for various geometries of the medium. 
We  define an effective accretion rate ${\dot m}_c \sim {\dot m}_b / \eta$ 
such that the cooling due to inverse Comptonization per gram of matter in the post-shock region 
with ${\dot m}_c$ is the same as the bremsstrahlung cooling with accretion rate ${\dot m}_b$.
Essentially, we use bremsstrahlung cooling with an accretion rate enhanced by a factor  $\eta$
whose value depends on $\tau_0$ (DLC91; CT95). This way, our basic 
questions, namely, whether the shock exhibits an oscillation in the presence
of cooling, and whether such oscillation shows quasi-periodicity are answered.
It so happens that the power density spectra (PDS) we obtained have similar 
characteristics as of the PDSs of light curves observed 
in black hole candidates with quasi-periodic oscillations (QPOs), so we shall use the terms QPVs and QPOs interchangeably. 
As discussed in MSC96, the cause of the
QPOs or QPVs is generic, namely, rough agreement between the cooling time and the infall time. So
we believe that the exact cooling process is immaterial -- as long as the two problems have a similar cooling
time-scale $E/{\dot E}$, where $E$ is the energy density and ${\dot E}$ is the cooling rate, 
they would produce the same oscillations. Hence PDS should continue to show QPVs 
even when more rigorous treatment of Comptonization and other coolings is included. In future,
we plan to consider Comptonization more rigorously.

Unlike MSC96, where symmetry around the equatorial plane was assumed
(and thus the simulation was carried out only in one quadrant), we consider both halves 
above and below the equatorial plane. In this way, allowance is made not only for radial 
oscillations of the shocks  but also for vertical oscillations.  Earlier, in a 
simulation of a non-shock flow it was observed that the outgoing winds could interact with the 
inflow and the interface might undulate (Molteni et al. 2001). So we expect that shocks 
should also have vertical oscillations in presence of winds. In the present simulations,
we observe that the shocks do oscillate both vertically and horizontally for a
range of inflow parameters, and the PDS of the light curve obtained by
our numerical simulation does have characteristics similar to the PDS of the observed light curves
of the black hole candidates. The PDS of a numerically simulated light 
curve has a flat top up to a break frequency, and an {\it  rms} 
amplitude in the range  of a few percent depending on the accretion rate of the flow.
Considering that an axisymmetric shock fundamentally separates a flow 
into a fast moving (lower density) supersonic and a slow moving 
(higher density) sub-sonic region, it should not be surprising that its oscillation
frequency will  generally lie at or near the break frequency in the power law.
Of course, shocks are not infinitesimally thin. Neither is it true that
they exhibit only a single mode of oscillation. Thus broad and often multiple peaks
are expected. This is precisely what we see. What is more, since the inner sonic point 
also separates the flow into two parts, namely, one with the disk-like behavior and the  one with free-fall
behavior, there is often a weak peak in the PDS at some hundreds of Hz, perhaps corresponding 
to oscillation of this inner region. We shall show these as well. 

In the next section, we present the basic setup of our numerical simulation.
In Sect. 3 we present some results of the simulation for galactic and extra-galactic 
black holes and analyze them to demonstrate shock oscillations. From the PDS of the 
`simulated' light curves, we show that there are sharp peaks at certain
frequencies only. The PDS seems to have the familiar 
flat top shape with a break. In Sect. 4, we discuss possible cause of the
behaviour of the shocks. Finally, in Sect. 5 we draw our conclusions. 

\section{Setup of the numerical simulations} 

We are interested in simulating the behavior of an axisymmetric, inviscid flow
in the presence of cooling effects. Unlike an ordinary star, a black
hole is capable of accreting matter even in the absence of viscosity 
when the pressure is large enough. However, if the specific angular 
momentum is less than the marginally stable value ($3.67 GM_1/c$ for 
a Schwarzschild black hole of mass $M_1$), even a cool, non-viscid
gas can enter a black hole without any problem. We chose the 
specific angular momentum of the flow $\lambda$ to be lower compared to this value
everywhere. This makes $\lambda < (G M_1 r)^{1/2}$, the Keplerian value, 
everywhere in the flow. We call such a flow a `sub-Keplerian' flow. The motivation for
studying such a flow stems from the fact that accretion processes 
into a black hole are necessarily transonic, and the flow has to be 
sub-Keplerian near the horizon. Furthermore, a flow can have a 
substantial amount of sub-Keplerian matter itself because of
accretion of winds from the companion. The wind of velocity $v_w \sim (2GM_2/R_2)^{1/2}$, 
having `zero' angular momentum with respect to the
mass-lossing star of mass $M_2$ and radius $R_2$ has an angular momentum of 
$l_w \lsim v_w R_a$, where $R_a \sim 10^{11}$cm is the radius of the accretion disk. For typical
values: $M_2\sim 2 \times 10^{33}$gm, $R_2\sim 10^{11}$cm and we get,
in units of $2GM_1/c$, $l_w \lsim 3.2$ with $M_1\sim 10M_\odot$. Hence,
our choice of $1.75$ which is averaged over the entire 
flow (thus could be much smaller than $l_w$) is justified. 
Recent observations do indicate the presence of both Keplerian 
and sub-Keplerian components (Smith, Heindl \& Swank, 2002) in half a dozen
black hole components.

It was shown by Chakrabarti (1989, 1996a) that in the large region 
of the parameter space spanned by the specific angular momentum $\lambda$ 
and energy ${\cal E}$ a stable standing shock may form in accretion whenever 
the Rankine-Hugoniot conditions are satisfied. Here, the flow first passes through the
outer saddle-type sonic point, then through a standing shock and finally through the
inner saddle-type sonic point. It was subsequently shown by MSC96 and
Ryu, Chakrabarti and Molteni (1997) that even when the parameters of the 
injected flow are outside this region oscillating shocks may form.
In the present paper we use Smoothed Particle Hydrodynamics (SPH) to do the 
simulations as in MSC96, but we use both halves of the flow 
to inject matter. The code has already been tested for its accuracy against 
theoretical solutions already (Chakrabarti \& Molteni, 1993, MSC96) and we do not repeat this here. 
We just want to recall that the `pseudo'-particles have been chosen 
to be `toroidal' in shape and they preserve angular momentum very 
accurately (Molteni, Ryu \& Chakrabarti, 1996). In MSC96, 
suggestion was made that {\it radial} oscillations of the shock may 
take place when the infall time in the post-shock region is comparable to the 
cooling time scale. Presently, we concentrate on the vast region of the parameter space 
where both the {\it radial and vertical} oscillations are present. 

In the context of galactic black holes one class of QPVs have been observed that are commonly 
known as Quasi-Periodic Oscillations or QPOs. Our solutions for QPVs are so generic that
we believe that QPVs should be observed even for super-massive black holes. 
Actually there are indications that QPOs in supermassive
black holes may have been observed (Halpern \& Marshall, 1996). Supermassive
black holes are believed to be fueled by low angular momentum matter generated from 
winds of nearby stars. So, a similar consideration 
as that for wind accreting galactic black holes will apply. As far as Comptonization 
is concerned, we use the following simple procedure: (a) We use an accretion rate ${\dot m}_b$
and run the code with bremsstrahlung cooling alone.  (b) We get the temperature 
of the electrons and the optical depth $\tau_b$ of the 'mean' size of the CENBOL region 
($X_s$ in Table 1). To get the corresponding  physical parameters (as if the Compton cooling were included
to begin with) (c) we assume that the injected soft
photons have energy $\sim 1$kev for the stellar mass black holes and $\sim 0.01$kev for  the 
massive black holes and using that (d) we compute the enhancement factor $\eta$ from DLC91 
for uniform spherical geometry for this injected photon and the electron temperature 
(assumed to be lower by a factor of $(m_p/m_e)^{1/2}$ than the proton temperature [e.g., Rees, 1984]) obtained
in each run. (e) We reduce the accretion 
rate ${\dot m}_b$ by this $\eta$ to get ${\dot m}_c={\dot m}_b/\eta$ and the optical
depth by $\tau_c=\tau_b/\eta$. For self-consistency, we ensure that $\tau_c$ thus obtained
is the same as that used while computing $\eta$ from DLC91. We also 
compute the Compton $y$ parameter to check if it has an acceptable value.
This procedure is re-formulates the problem of inclusion of the exact cooling
processes into the case where a simplified cooling effect
may be used. Since we are interested in the integrated energy loss due to cooling
and not the spectral properties, this procedure should give reasonable answer.

Table 1 lists the parameters for the Model runs 
we report in this paper. There are basically two groups of 
input. In Group A, we consider the black hole to be super-massive ($M=10^8M_\odot$) 
and in Group B, we consider a stellar mass black hole ($M=10M_\odot$). Different
runs are characterized by the injected matter density (gm/cc) at the outer boundary 
of the numerical grid given in the second column. In all the model runs, we choose 
the following parameters: the index $\gamma$ in the polytropic relation $P \propto \rho^{\gamma}$ 
($P$ is the isotropic pressure and $\rho$ is the gas density) is $5/3$, the outer 
boundary $r_{out}=50$, the specific angular momentum $\lambda=1.75$ (in units of $2GM/c$), 
injected radial velocity $\vartheta_r=0.126$, the sound velocity $a=0.04$ 
and the vertical height $h=15$. The vertical height at 
the outer boundary is so chosen that the flow remains in hydrostatic
equilibrium in the vertical direction at the point of injection 
$h \sim a r_{out}^{3/2}$. We measure all the 
distances in units of the Schwarzschild radius of the black hole 
$r_g=2GM_{BH}/c^2$, all the velocities in units of the velocity of 
light $c$ and all the masses in units of $M_{BH}$, the mass of the 
black hole. $c$ and $G$ are the velocity of light and the
gravitational constant respectively. Note that the 
marginally stable (lowest possible angular momentum with a stable Keplerian
orbit) angular momentum is $1.83$ in these units. 
In the third column we give the enhancement factor $\eta$ that we obtained using DLC91 for our run
and in the fourth column we provide ${\dot m}_c$ in units of ${\dot M_{Edd}}$,
the Eddington rate, the effective accretion rate
in the presence of Comptonization. In the fifth column we give the Compton $y$ parameters.
We assume soft injected photon of energy $h\nu_{inj} \sim 1$keV for galactic  black holes, and $0.01$keV 
for supermassive black holes in order to calculate $\eta$. In the sixth column,
we give the average observed location of the shock.
In the seventh column, we give the average number of `pseudo'-particles present in 
the disk in each run. In the eight column we present the QPV(QPO) frequency 
in Hz as obtained from PDS. In the ninth column, the inverse of the infall timescale
$t_i \sim X_s^{3/2}$ corrected for a constant compression ratio $R$ ($\sim 4$ for a
strong shock) is provided as an indication of the expected oscillation time scale.
In the tenth column, the ratio of the quantities in the eighth and ninth column
is given. In the eleventh column, the $rms$ amplitude in percentage is given,
we shall discuss in the next section. As in MSC96, 
we choose the Paczy\'nski-Wiita (1980) pseudo-Newtonian potential 
to describe the space-time around the Schwarzschild black hole. 
We believe that the result would remain unchanged if a true Schwarzschild
geometry were used, since the transonic properties do not change 
very much. However, if a Kerr geometry were used, we expect the locations of the
inner sonic points and shocks to move inward (Chakrabarti, 1996b) giving rise to higher QPO frequencies.

\centerline{ TABLE 1: Inputs and extracted parameters for the model runs}

\begin{center}
\begin{tabular}{lllllllllllll}
\hline
\hline
Run &$\rho_{inj}$ & $\eta$ & ${\dot m}_c$ & y & ${\bar X}_s$ & ${\bar N}$ & $\nu_{QPO}$ & $(Rt_i)^{-1}$ & $\frac{\nu_{QPO}}{(Rt_i)^{-1}}$ &  $R_{\it rms}$  \\
 & (gm/cc) &  &   &  & & & (Hz) & & & ($\%$) \\
\hline
\hline
A.1 & 0.95e-14 & 90 & 0.27 & 5.1& 16 & 22250 & 1.83e-6 & 3.9e-6 & 0.46  &  11.2 \\
&  & & &  & & & 9.15e-8 &  & &  \\
A.2 & 1.26e-14 & 142 & 0.22 & 6.6& 13 & 19250 &  2.06e-6 & 5.3e-6 & 0.39 & 11.2  \\
& &  & & &  &  & 5.34e-7 & & &   \\
A.3 & 3e-14 & 208&  0.37  & 8.3 & 6  & 12875 &  1.37e-6& 1.7e-5 & 0.23  &   11.2 \\
& &  & &  & & & 3.36e-7 & & &  \\
A.4 & 5e-14 & 331 & 0.39 & 9.4 & 5  & 10500 & 6.88e-7 & 2.2e-5 & 0.03 & 3.3 \\
&  & & & & & &  3.57e-7  & & &  \\
\hline
B.1  & 3.6e-10 & 1.8& 0.05 & 0.4& 15.8  & 21950  & 19.34  & 40 & 0.48 & 3.5\\
B.2   & 4.5e-10 & 1.5& 0.08 & 0.5 & 16.3 &22050  & 19.45 & 38.5  & 0.51 &  4.0 \\
B.3   & 4.5e-8 & 29.4 &0.39 & 11.8 & 24  & 33500  & 10.2& 21.3  & 0.48 &  14 \\
 & & &  & & & & 2.13 &  & & \\
B.4   & 4.5e-7 & 134 & 0.85 &43.4 &5 & 10825  &  8.32  & 223  & 0.04 & 3.7 \\
 & & & & &  & & 3.58 &  & &  \\
\hline
\hline
\end{tabular}
\end{center}

One of the important conditions of a standing shock
is that the thermal pressure ($P$) plus ram pressure ($\rho \vartheta_r^2$) should be 
continuous across the shock in the steady state, i.e.,
$$
P_- + \rho_- v_-^2 =P_+ + \rho_+ v_+^2 .
\eqno{(2)}
$$
Here, the subscripts $-$ and $+$ denote quantities before and after the shock 
and $v$ denotes the radial velocity of the flow as measured in local rotating frames.
When we increase the density of the gas, the cooling rate is increased. 
The thermal pressure is thus decreased, especially in the post-shock region,
causing the shock to move closer to the black hole.
In our simulations we generally see this 
trend in Group A runs (Table 1). Along with the shock location, there is a systematic 
decrease in frequency as well since the flow becomes cooler and the timescale of 
cooling itself is increased. 

In Group B Runs, within our parameter range, occasionally there were stronger 
shocks which oscillated radially, and the interaction between the outflow and the 
inflow produces vertical oscillations as well. Strong convection 
and turbulences were seen in the post-shock region which contributed to pushing the 
shock upstream. As a result, the shock location was not seen to change 
monotonically with accretion rate. The location is clearly determined by 
several effects: centrifugal force, thermal pressure, ram pressure, turbulent
pressure  and the cooling rate. The nature of the dependence has been discussed in earlier
work in detail (Chakrabarti, 1989; Chakrabarti \& Molteni, 1995; Molteni, Lanzafame \& 
Chakrabarti, 1994; MSC96) and we do not repeat this here. We will just recall that 
the shock location is independent of the location of the 
numerical boundary when the fundamental quantities such as specific angular 
momentum, accretion rate and specific energy at a given point (say, at the 
inner sonic point) are kept fixed.  Cases B.1 and B.2 have very small 
Compton parameters and negligible Comptonization. Cases B.3 and B.4 are 
presented for academic purposes, just to indicate the effects of excess cooling. 
The Compton parameters are very large in these two cases. 

For reference, we may add that the light crossing times
of the horizon for the two classes of black hole are $r_g/c \sim 10^3$s 
and $10^{-4}$s respectively. Since a steady shock (in the Schwarzschild geometry) may form anywhere between 
$X_s\sim 10$ to $100r_g$, the QPO time-period, if assumed to be 
related to the infall time $t \propto X_s^{3/2}$, may be close 
to $10^5$ to $10^8$s for super-massive black holes, and close to $10^{-2}$s 
to $10$s for stellar mass black holes. Since we chose the outer boundary of 
the simulation to be at $r_{out}=50$, the infall time is roughly $r_{out}^{3/2} 
\sim 350$ in units of $r_g/c$. In order to trust the results of our simulations, 
we ran the code for a duration several hundred times of this timescale (typically, $T_{run}\sim
50,000-60,000$ or more). For a black hole of $M=10M_\odot$, this time corresponds to only $5-6$s
of real time. 

\section{Results}

First, we show examples of shock formation which exhibit radial and/or 
vertical oscillations.  Fig. 1(a-c) shows the locations of the SPH 
particles (dots) along with the velocity vectors (arrows) of 
every fifth particle for clarity. This case corresponds to 
case A.2. The matter distribution is shown at three different times (in
units of $r_g/c$) illustrating the vertical and radial motions of 
the shock which oscillates  between $11$ to $15$ Schwarzschild radii, mostly 
staying close to $r_s\sim 13r_g$. Some matter can be seen bouncing back
from the centrifugal barrier (boundary between the empty region along the vertical axis and the disk/jet 
matter) near the axis and forming giant vortices which interact with the 
inflow. These vortices push the post-shock region ($r\lsim 15r_g$) 
alternately into the upper and the lower halves.

\begin {figure}
\vbox{
\vskip -3.0cm
\hskip 0.0cm
\centerline{
\psfig{figure=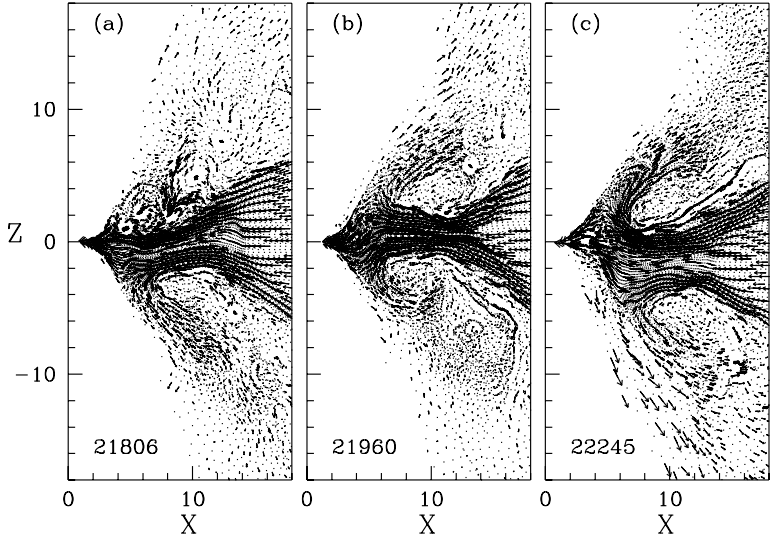,height=13truecm,width=13truecm}}}
\vspace{0.0cm}
\caption[] { Snapshots of simulations of accretion disks around a $10^8M_\odot$ black hole
by Smoothed Particle Hydrodynamics. The dots are particle locations and arrows are
drawn for every fifth particle for clarity. Time (in units of $r_g/c$) is marked
in each box. Note the vertical as well as radial oscillation of the accretion shock wave
located at $\sim 13r_g$.}
\end{figure}

In Fig. 2(a-d) we show the variation of the light curve in runs A.1-A.4 (marked 
by a-d in the Figure). The light curves are  the variations of the
luminosity of bremsstrahlung radiation (in arbitrary units) for all the
matter within $r=50$ flow as a function of time (in seconds). 
As the density is increased, the average thermal energy gradually goes down
due to the presence of enhanced cooling and thus the luminosity also
goes down. The average location of the shock decreases (MSC96, Table 1).
The average number of pseudo-particles (Table 1) as well as the variation of the 
number of these particles also go down. As a result, with  
the increase in cooling loss, the light curves are 
found to be less noisy, and the amplitude of fluctuation is found to be lower. 

\begin {figure}
\vbox{
\vskip -0.0cm
\hskip 0.0cm
\centerline{
\psfig{figure=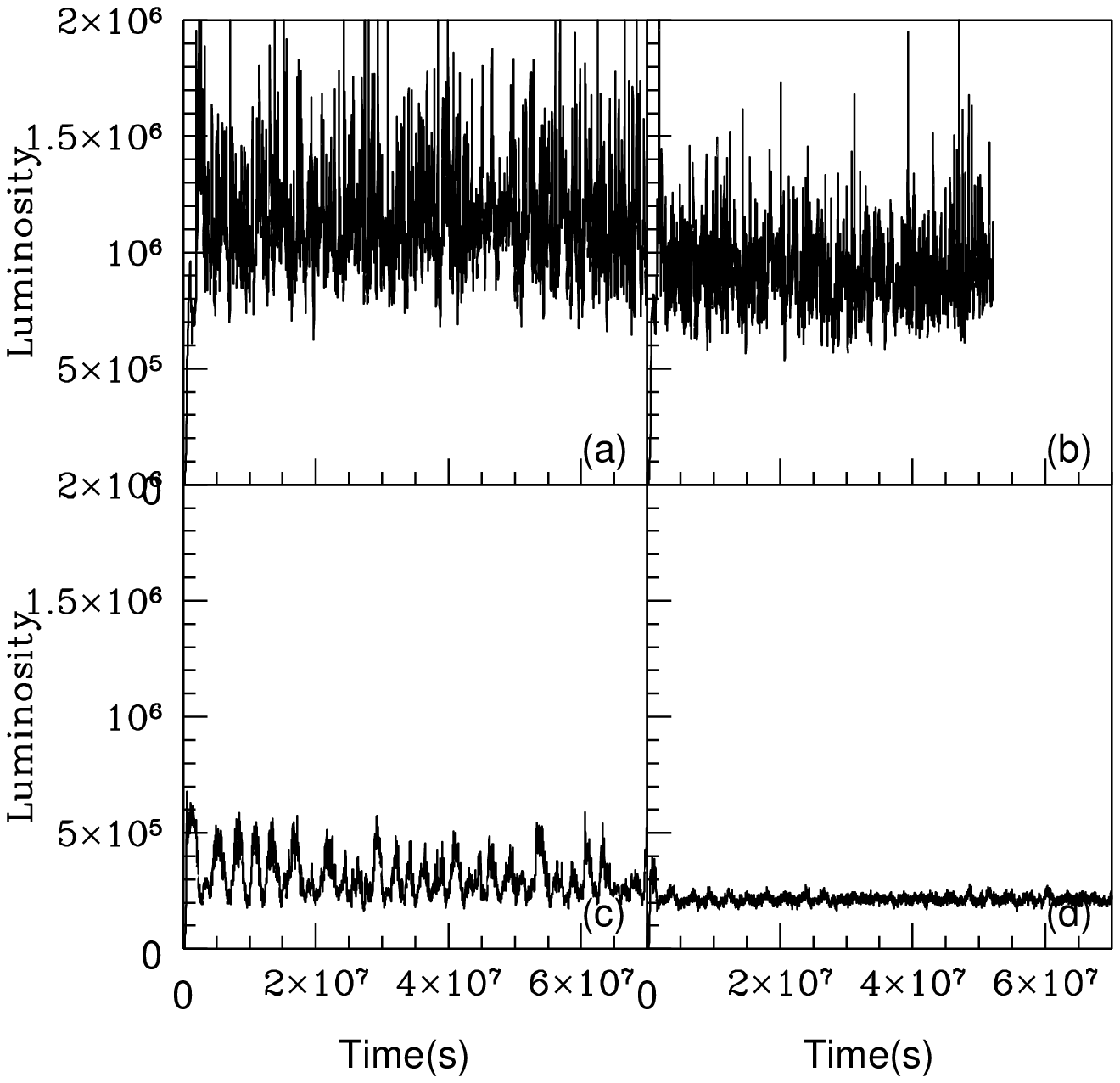,height=13truecm,width=13truecm}}}
\vspace{-2.0cm}
\noindent {\small {\bf Fig. 2(a-d)}}:
Total bremsstrahlung luminosity of the accretion flow
(in arbitrary units) as a function of time (in seconds).
The injected densities are (a) $\rho_{inj}=0.95\times 10^{-14}$gm/s,
(b) $1.265 \times 10^{-14}$gm/s, (c) $3 \times 10^{-14}$gm/s and
(d) $5 \times 10^{-14}$gm/s respectively. The disk becomes
cooler with increasing accretion rate due to bremsstrahlung
energy loss.
\end{figure}

In Fig. 3(a-d) we show the results of the Fast Fourier Transform of these light curves using FTOOLS
provided by NASA, the same software package that is used to 
analyze observational results. We find the remarkable result of familiar
power density spectra (PDS) with a QPV near the break frequency.
The QPV frequencies, indicated by the arrows, are presented in Table 1. 
We find that, perhaps due to the
presence of both the vertical and radial oscillations, there are 
two QPVs; the frequency of the stronger one gradually increases with 
the cooling, while that of the weaker one gradually decreases. 
The occurrence of QPV frequencies  at or near the break frequency of the PDS
can be understood by the following: according
to our solution, QPVs occur due to oscillations of the shock waves that separate
two `phases' of the flow --- the pre-shock flow is supersonic and weakly radiating,
while the post-shock flow is subsonic and strongly radiating. The former 
region of the flow produces the `flat-top' region in the PDS, while the latter region produces
a PDS with a slope $\sim -2$. The rms amplitude of the flat top
region up to the break frequency is shown in Table 1. It has generally similar value at around $11.2$\% except 
in (d) where it is only around $3.3$\%. These rms values are similar to the observed results from black hole 
candidates as well. In Fig. 3(a-b), where the cooling is weaker, the QPVs are less prominent
and we had to search the locally significant peaks of the PDS for them. As the cooling becomes stronger, 
the peak becomes stronger. 

\begin {figure}
\vbox{
\vskip -0.0cm
\hskip 0.0cm
\centerline{
\psfig{figure=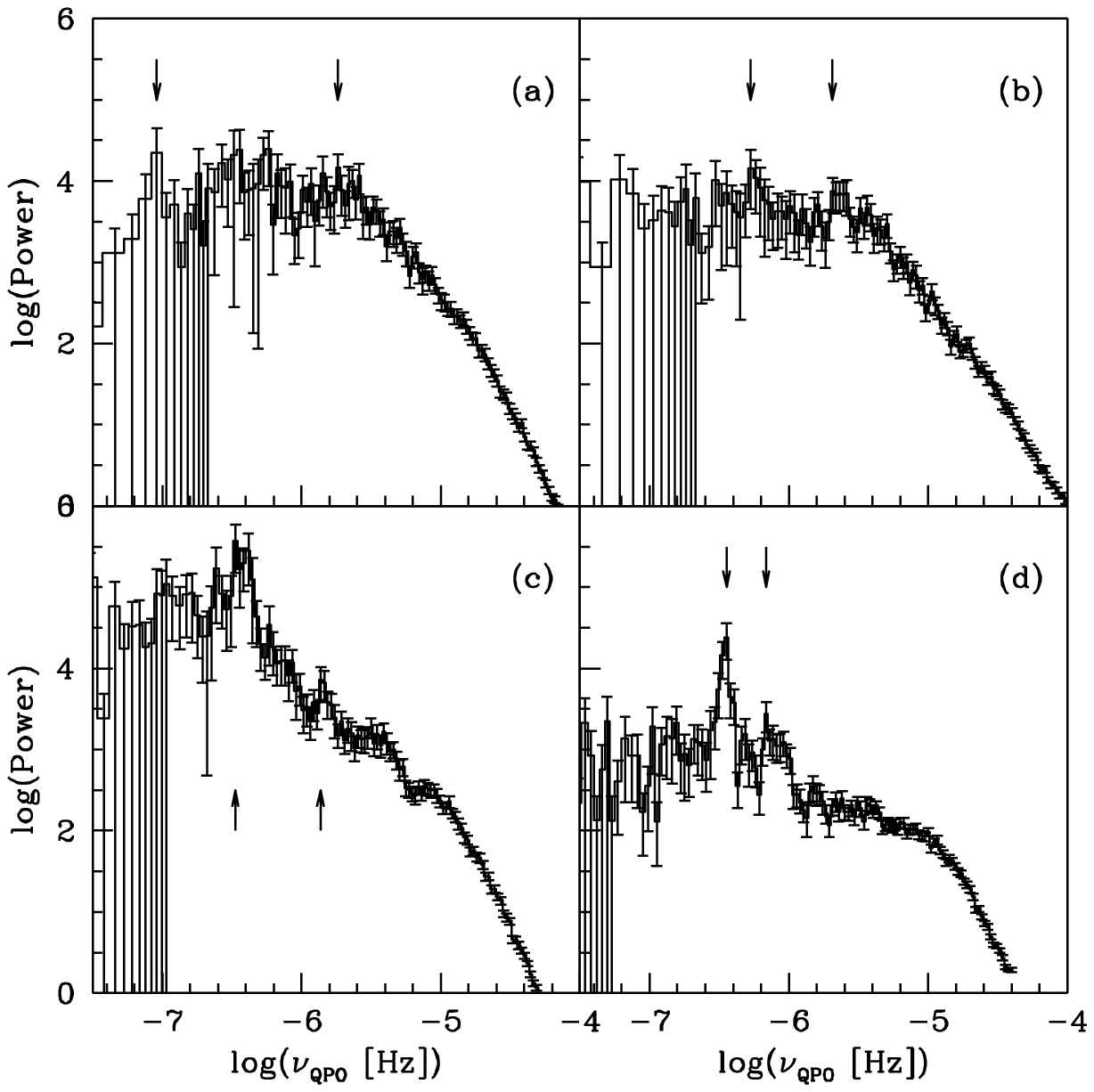,height=13truecm,width=13truecm}}}
\vspace{-2.0cm}
\noindent {\small {\bf Fig. 3(a-d)}}: Power Density Spectra (PDS) of the four cases shown in Fig. 2.
Quasi Periodic Variation frequencies (denoted by $\nu_{QPO}$) can be
seen with timescales of hours to weeks. QPV peaks
(marked with arrows) are located near break frequencies with flat top
behavior at low frequency and power-law behavior
at high frequency. See Table 1 for other properties.
\end{figure}

In Fig. 4(a-d), we present similar results as in Fig. 2 for stellar mass black holes
(runs B.1 - B.4 of Table 1). Here, too, the disk becomes cooler with increasing density 
and the light curve changes its character according to whether the shocks are strong or not. 
In the beginning of the simulation some transient behaviour is seen till the 
disk settles down within a second in real time.
In Fig. 4(c), the parameters are such that a strong shock forms and it heats up the
disk while for the parameters in Fig. 4(d) a very weak shock forms. Shock oscillation
produces a large amplitude noise in the B.3 case (Fig. 4c). In Fig. 5(a-d),
the PDS of these four runs are shown. The transient region has been excluded 
while doing the FFT. As in Figs. 3(a-d), here too QPVs are
observed. For small black holes QPVs occur at a few Hz and the QPV is located at or 
near the break frequency. Thus the characteristics are similar to those of QPOs observed from 
black hole candidates. Radial and vertical oscillations often produce 
multiple peaks in this case as well. Other details can be seen in Table 1.
In Fig. 5c the peak is not very strong, perhaps due to the large amplitude noise
that is produced in this case (See, Fig. 4c). 

\begin {figure}
\vbox{
\vskip -0.0cm
\hskip 0.0cm
\centerline{
\psfig{figure=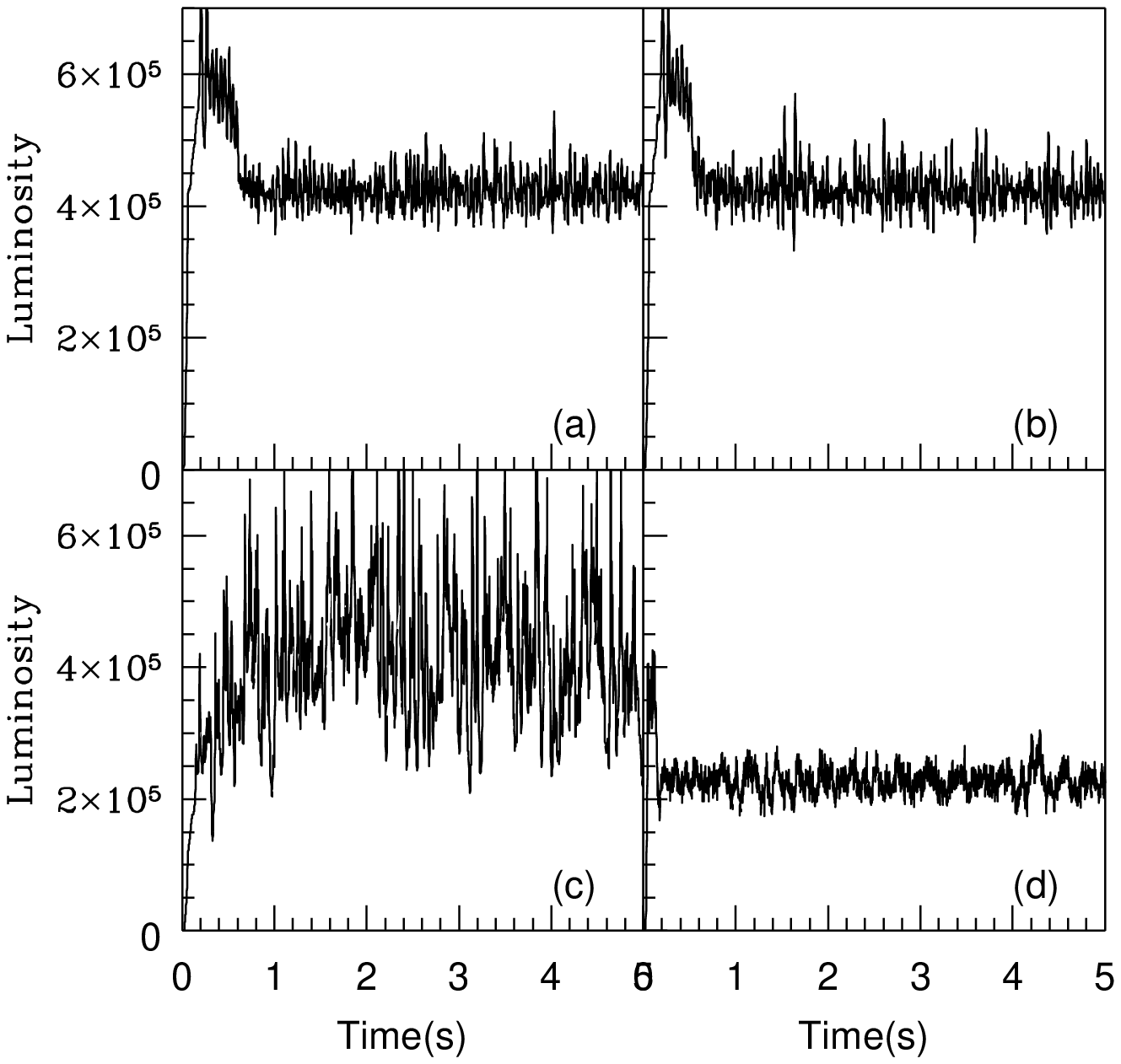,height=13truecm,width=13truecm}}}
\vspace{-2.0cm}
\noindent {\small {\bf Fig. 4(a-d)}}: Total bremsstrahlung luminosity of the accretion flow
(in arbitrary units) as a function of time (in seconds).
The injected densities are (a) $\rho_{inj}=3.6 \times 10^{-10}$gm/s,
(b) $4.5\times 10^{-10}$gm/s, (c) $4.5 \times 10^{-8}$gm/s and
(d) $4.5 \times 10^{-7}$gm/s respectively. The disk becomes
cooler with increasing accretion rate due to bremsstrahlung
energy loss.
\end{figure}

Occasionally, peaks at a very high frequency could be seen, but these peaks are
found to be transient. If the entire light curve is broken into smaller
pieces, weak peaks at frequencies $\sim 100-300$Hz can be seen, though they
do not persist. One example of this is shown in Fig. 6(a) where the PDS of the 
average thermal energy up to $T=36000$ is plotted for the case B.4. In Fig. 6(b) and (c) 
we show the flow pattern at two times which are separated by only $18$ units
($0.0018$s). The shock located at $\sim 6r_g$ shows two distinctly different
shapes at these two times, bent up and bent down. Oscillation of a density enhancement at $\sim 6r_g$ 
would have a time period of $\sim 6^{3/2} \times 10^{-4} R \sim 0.0015 R $s, where $R$, the compression
ratio, is $\sim 2 $ for a weak shock. The corresponding frequency is $333$Hz. The 
weak peak at $306\pm 7.93$Hz (Fig. 6a) may be due to this oscillation.
When data for much longer time is analyzed, this weak peak disappears completely.

\begin {figure}
\vbox{
\vskip -0.0cm
\hskip 0.0cm
\centerline{
\psfig{figure=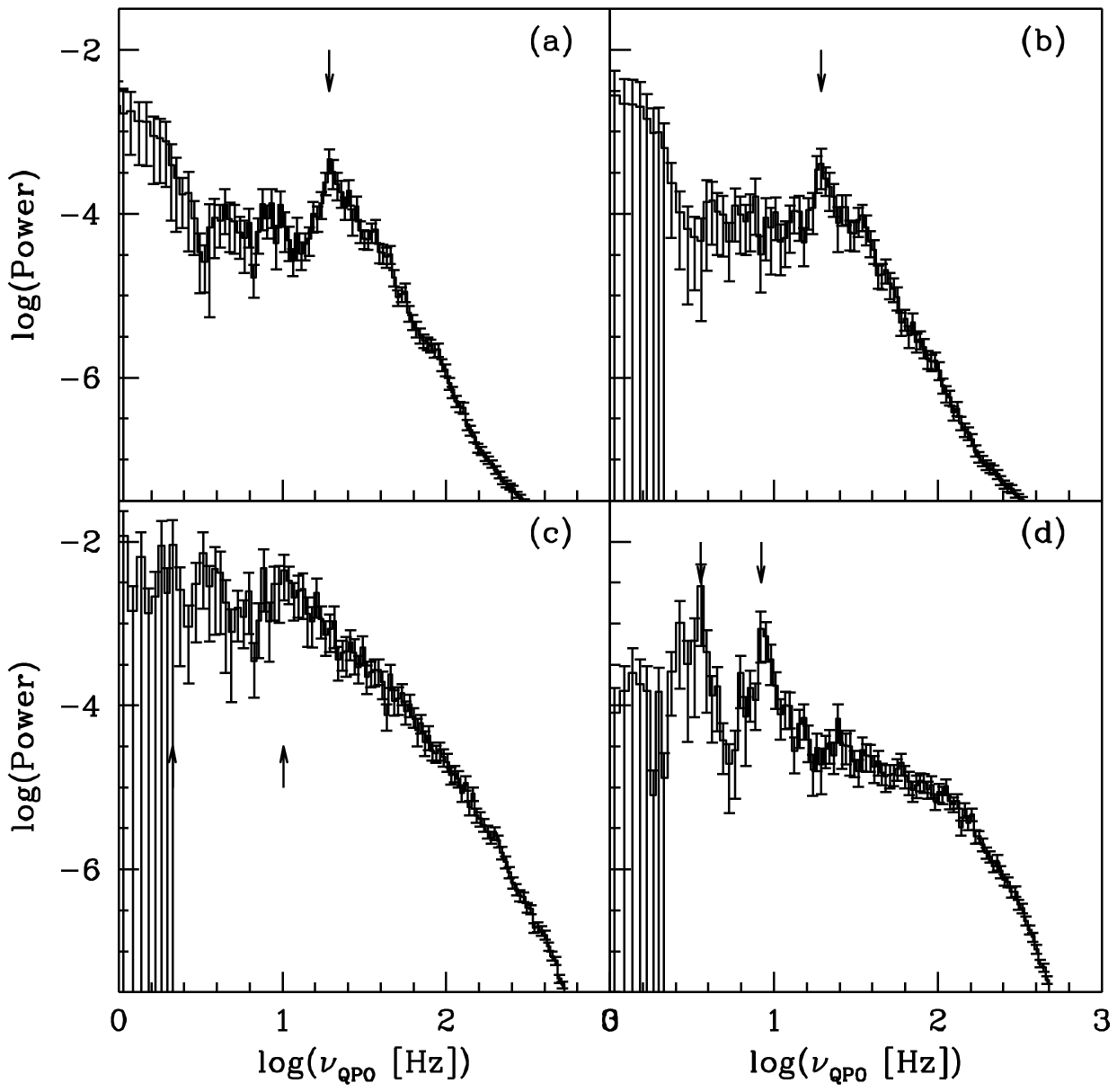,height=13truecm,width=13truecm}}}
\vspace{-2.0cm}
\noindent {\small {\bf Fig. 5(a-d)}}: 
Power Density Spectra (PDS) of the four cases shown in Fig. 4.
Quasi Periodic Variation frequencies (denoted by $\nu_{QPO}$) can be
seen in the range of $2-10$Hz and also
close to several hundred Hz. QPV peaks (marked with arrows)
are located near break frequencies with flat top
behavior at low frequency and power-law behavior
at high frequency. See Table 1 for other properties.
\end{figure}

\begin {figure}
\vbox{
\vskip -0.0cm
\hskip 0.0cm
\centerline{
\psfig{figure=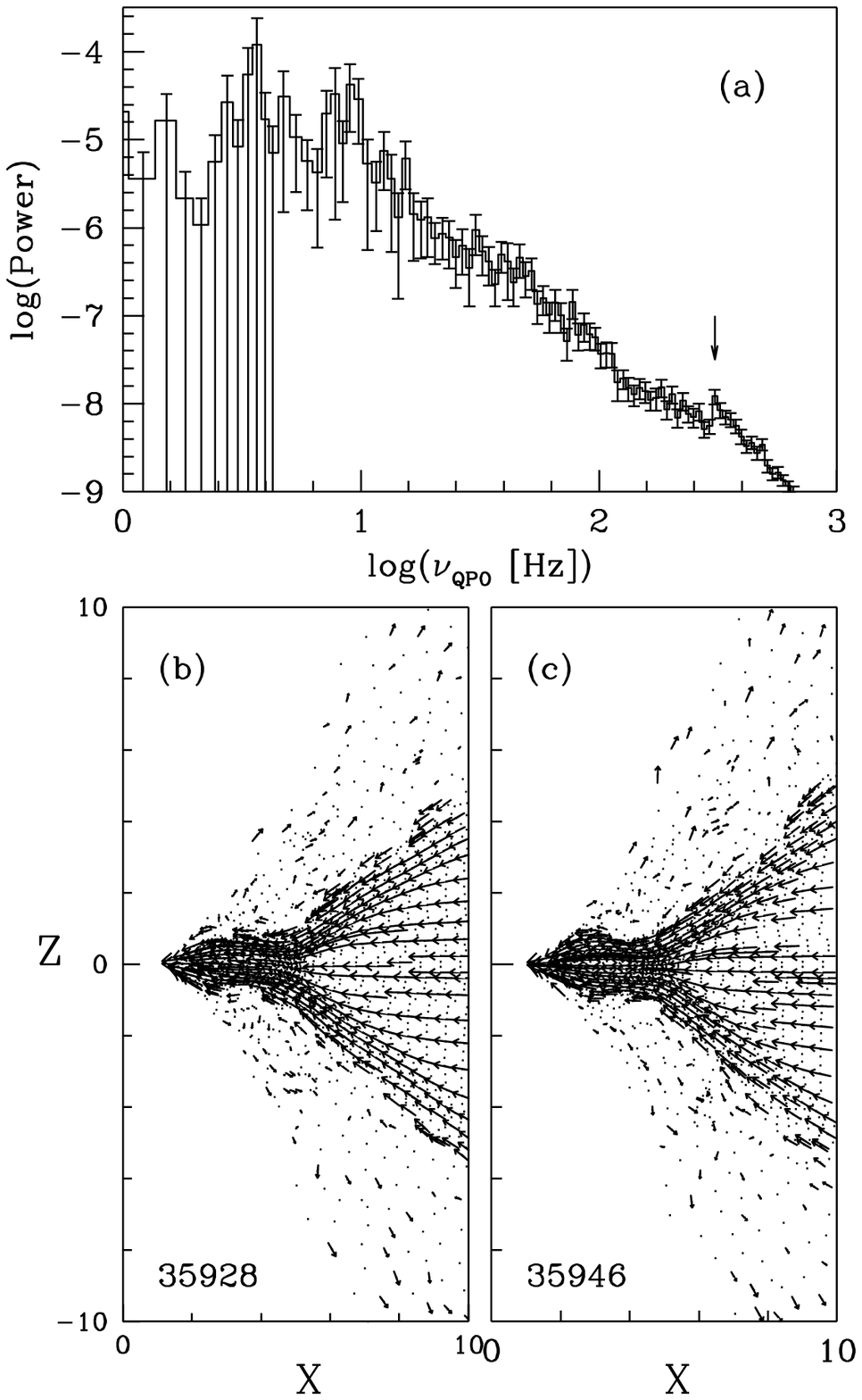,height=13truecm,width=13truecm}}}
\vspace{-0.0cm}
\noindent {\small {\bf Fig. 6(a-c)}}: 
(a) Power density spectrum of the average thermal energy for case B.4 with time up to
3.6s. Snapshots of simulations at closely separated times (marked) are
shown in (b-c). Only the inner $10$ Schwarzschild radii are shown.
The dots are particle locations and arrows are
drawn for every fifth particle for clarity. Time (in units of $r_g/c$) is marked
in each box. The oscillation of this region causes QPV at $\sim 300$Hz [marked with an arrow
in (a)].
\end{figure}

\section{The possible cause of shock behaviour and the resulting QPV}

A shock remains steady, or `standing' if Eq. (2) along with the 
specific energy and angular momentum flux remain continuous across it.
These conditions are known as the Rankine-Hugoniot 
conditions (Landau \& Lifshitz, 1959). However, in the presence 
of an efficient energy extraction mechanism from the disk, whether through thermal/non-thermal radiation 
or through higher entropy flows along the shocked winds and outflows, the shock  should
move closer to the black hole as the pressure in the post-shock region ($P_+$) drops.
This is what is seen  (see, Table 1), except in case B.3 where the turbulence was
too strong to push the shock outward. 

What about the oscillation time scale?  We note that the cooling time 
scale $ t_c\sim {\cal E}/{\dot {\cal E}}$ (where ${\cal E}$ is the 
specific energy and a `dot' indicates its time derivative) competes 
with the infall time scale $t_i \sim X_s/v$. MSC96 argued that 
when these two time scales agree roughly, i.e., roughly when the resonance takes place,
the shocks may start oscillating. The infall time $t_i$ is not easy to compute,
however. CM00 argued that the infall velocity need not be free-fall 
velocity $v\propto 1/x^{1/2}$ as the post-shock flow is reduced at the shock by at least 
a factor of $R$, the compression ratio. Furthermore, the velocity is 
highly modified due to turbulent pressure and angular motion and could be very slowly varying
in between the shock and the inner sonic point (CM00; Chakrabarti, Ryu \& Samanta, in preparation). 
A third and perhaps the most important parameter in predicting the time scale
is the rate of outflow from the post-shock region and the extent to which 
the total mass of the disk oscillates. These two cooling mechanisms, namely, through radiation loss
and through outflow, are acting in opposite directions -- enhanced thermal cooling
reduces the pressure gradient force, thereby reducing the wind rate. 
The QPV timescale is expected to be close to  
$t_{QPV}=\nu_{QPV}^{-1}\sim t_c \sim Rt_i$. In Table 1, $\nu_{QPV}/(Rt_i)^{-1}$
is presented. This ratio seems to be closer to $0.5$ (justifying a {\it near} resonance condition)
except when the cooling is very strong (runs A.4 and B.4) when the
ratio drastically goes down. In these latter types of runs the shock moves very close to the
black hole and becomes very weak. The QPV could be due to the domination of the
interaction of the winds with the inflow. The two QPV frequencies
often seen could be due to the two types of cooling.

\section{Concluding remarks}

In this paper, we presented the discovery of Quasi-Periodic 
Variations of emitted radiation  through extensive time-dependent 
numerical simulations. In earlier studies, such as in MSC96 and CM00, it was
suggested that shock oscillations might contribute to the observed QPOs. 
In the present work, we studied shock oscillations in low-angular-momentum 
axisymmetric flows around stellar and super-massive black holes for a range of accretion rates.
Unlike MSC96, we used fully two dimensional inflow (upper and lower halves)
and demonstrated that the shocks participate in vertical and radial oscillations. 
Using PDS of light curves obtained from time dependent numerical 
simulations of sub-Keplerian flows, we demonstrated that the  QPVs 
could be the cause of the observed QPOs.
We showed that the computed light curve produces power density spectra with 
characteristic QPVs located at or near the break frequency which separates the flat-top 
type and power-law type PDSs. For stellar black holes, we see QPVs at 
reasonable frequencies ($2-20$Hz). Very often high frequency QPVs 
(at around $100-300$Hz) could be observed. We demonstrated that
this could be due to oscillation at the very inner edge where the flow becomes supersonic.
For Active Galaxies and Quasars, QPVs at very low frequencies ($\sim 10^{-7}-10^{-6}$ Hz)
are observed. Their properties are similar to those of QPOs observed in these candidates.

Based on our simulations we propose that the QPOs observed in black hole 
candidates are genuinely related to the accretion shock oscillations. 
Variation of QPO frequencies from time to time in the same object are thus
manifestations of the variation of the Keplerian and sub-Keplerian accretion rates. 
In our work, we assumed a simple  Paczy\'nski-Wiita potential 
to study the shock properties around Schwarzschild black holes. Based on our experience
with transonic flows, we are confident that none of the properties would be affected 
if the simulations were carried out in true Schwarzschild geometry. However, if a rotating
black hole is used, the shocks will form much closer to the black hole (Chakrabarti, 1996b), resulting in
higher QPV frequencies. 

SKC thanks the support of Indian Space Research Organization for a RESPOND project. 
Discussions with Prof. A.R. Rao (TIFR) and Mr. A. Nandi (SNBNCBS) are also acknowledged. 

{}

\end{document}